\documentclass[3p]{elsarticle}
\usepackage{amsmath,amssymb,multirow,epsfig,bm,subfigure}
\usepackage{tabularx}

\newcommand{\beq}{\begin{equation}}
\newcommand{\eeq}{\end{equation}}
\newcommand{\bea}{\begin{eqnarray}}
\newcommand{\eea}{\end{eqnarray}}

\begin{document}
\title{Preconditioning the non-relativistic many-fermion problem}

\author[CHI]{Timour Ten}
\ead{tten1@uic.edu}

\author[OSU]{Joaqu\'{\i}n E. Drut\corref{cor1}}
\ead{jdrut@mps.ohio-state.edu}

\author[AAU]{Timo A. L\"ahde}
\ead{timo.lahde@tkk.fi}

\address[CHI]{Department of Physics, University of Illinois, Chicago, IL 60607--7059, USA}

\address[OSU]{Department of Physics, The Ohio State University, Columbus, OH 43210--1117, USA}

\address[AAU]{Helsinki Institute of Physics and Department of Applied Physics,
Aalto University, FI-02150 Espoo, Finland}

\cortext[cor1]{Corresponding author}

\date {\today}

\begin{abstract}
Preconditioning is at the core of modern many-fermion Monte Carlo algorithms, such as Hybrid Monte Carlo, where the repeated solution 
of a linear problem involving an ill-conditioned matrix is needed. We report on a performance comparison of three preconditioning strategies, 
namely Chebyshev polynomials, strong-coupling approximation and weak-coupling expansion. We use conjugate gradient (CG) on 
the normal equations as well as stabilized biconjugate gradient (BiCGStab) as solvers and focus on the fermion matrix of the unitary 
Fermi gas. Our results indicate that BiCGStab is by far the most efficient strategy, both in terms of the number of iterations and 
matrix-vector operations.
\end{abstract}
\maketitle

%%%%%%%%%%%%%%%%%%%%%%%%%%%%%

\section{Introduction}

The Hybrid Monte Carlo (HMC)~\cite{HMC_Duane,HMC_Gottlieb} algorithm provides an efficient way of calculating the properties of 
many-body Fermi systems. HMC achieves high efficiency by performing global updates by means of Molecular Dynamics~(MD) 
trajectories that are integrated with a finite time step. At each step, the solution of a linear problem
\beq
\label{MTMLinearProblem}
M^\dagger_{} M x = b
\eeq
is needed. In this problem, $M$ is a large but typically sparse matrix which provides a real-space representation of the 
fermion operator that defines the dynamics of the system. While $M$ is typically neither symmetric nor positive definite, it
is always possible to consider solving Eq.~(\ref{MTMLinearProblem}) via conjugate gradient (CG) iteration, as $M^\dagger M$ 
satisfies these properties~\cite{CG}. However, as CG is a Krylov-type method based of successive 
matrix-vector~(MV) operations using $M^\dagger M$, the efficiency of CG depends critically on the condition number of that matrix. 
Unfortunately, in most cases of interest $M^\dagger M$ may be extremely ill-conditioned, leading to a rapid growth of the number 
of CG iterations required to reach a preset convergence criterion. Such a situation manifests itself in many problems at low enough 
temperatures or strong enough couplings. Apart from the obvious disadvantages of a rapidly growing computational 
cost, any iterative method may eventually become unstable if the number of iterations grows without bounds. For these 
reasons, CG and similar Krylov methods are preferentially used together with a suitable preconditioner. It should be 
pointed out that while direct methods such as LU~decomposition avoid the problems related to ill-conditioning, problems
of physical interest are typically large enough such that the storage of $M$ is not feasible. It is thus desirable 
to resort to ``matrix-free" methods based on MV operations only, which require significantly less memory and scale much 
more favorably with system size.

Whenever $M$ is ill-conditioned, the problem is much more severe for $M_{}^\dagger M$. An attractive option is then to perform the
solve the problem in two steps, by considering the linear systems given by
\beq
M^\dagger_{} y = b, \quad\quad M x = y.
\label{MLinearProblem}
\eeq
However, since $M$ is typically neither symmetric nor positive definite, these equations require more sophisticated 
algorithms such as biconjugate gradient (BiCG)~\cite{BiCG} or stabilized biconjugate gradient 
(BiCGStab)~\cite{vanderVorst}, which is a modern development of the former. One is thus dealing with a much less 
ill-conditioned problem, at the price of solving two successive problems. Nevertheless, preconditioning remains 
instrumental in accelerating and stabilizing the solution process. In its simplest form, preconditioning amounts to 
considering the linear problem
\beq
\label{PMTMLinearProblem}
P M^\dagger_{} M x = P b,
\eeq
where $P$ represents an approximation to $(M^\dagger_{} M)^{-1}_{}$. A similar procedure applies to Eq.~(\ref{MLinearProblem}) given 
approximations to the inverses of $M$ and $M^\dagger$. The procedure in Eq.~(\ref{PMTMLinearProblem}) is referred to as 
``left preconditioning''. So-called ``right preconditioning'' is also possible by insertion of $P$ between $M^\dagger_{} M$ and $x$, 
followed by multiplication with $P$ once the solution has been obtained. In this work, we shall be mainly concerned with 
the former approach.

While the choice of $P$ is in principle arbitrary, preconditioning is only useful if the solution of 
Eq.~(\ref{PMTMLinearProblem}) to a given accuracy requires fewer CG iterations than the solution of 
Eq.~(\ref{MTMLinearProblem}). Moreover, as iterative methods tend to accumulate roundoff error, an iterative solution 
may not even be possible without preconditioning. In such cases preconditioning also serves to stabilize the algorithm 
by enabling convergence. Two obvious extreme choices are those of setting $P = I$ and setting $P = (M^\dagger_{} M)^{-1}$, however 
no benefits are achieved in either case, as the first one amounts to no preconditioning at all and the second one 
requires the solution of the original problem in order to determine $P$. Therefore, what one seeks is compromise between 
the computational simplicity of the first option and the effectiveness of the second one.

Over the years, many preconditioning strategies were developed and tested on a wide range of problems (see {\it e.g.} Chapter 11 in 
Ref.~\cite{BroydenVespucci}, or Chapters 9 and 10 in Ref.~\cite{Saad}). In general, preconditioners specifically 
designed for a particular problem tend to be superior to generic approaches that build in no knowledge about the physics 
of the system in question. With this in mind, we will explore several approaches to preconditioning which combine some knowledge of 
the features of $M$, $M^\dagger_{}$ and $M^\dagger_{} M$, in order to find a strategy which is both computationally efficient and effective 
in accelerating the convergence of the linear problem at hand.

In this work, we are mainly concerned with problems that arise in the context of HMC simulations of 
non-relativistic, (3+1)-dimensional many-fermion systems. The HMC approach forms a central part of state-of-the-art 
Lattice QCD calculations~\cite{LQCDReview}, which involve a relativistic system of fermions. The popularity of HMC stems from the great 
algorithmic efforts of the Lattice QCD community to enable simulations on large lattices. On the other hand, HMC remains 
little known in other areas such as condensed-matter physics~\cite{CondMatt} and nuclear structure~\cite{Leeetal}, where many of the 
problems are non-relativistic and determinantal Monte Carlo (DMC)~\cite{DMC} remains the method of choice. The 
most serious drawback of DMC is the poor scaling of the computation time with the lattice volume $V$, namely $\sim 
V^3_{}$. In contrast, the use of global updates and iterative solvers enables the HMC algorithm to scale as
$\sim V^{5/4}_{}$. In spite of this obvious advantage, the widespread use of HMC implementations for non-relativistic 
fermions has been hampered by problems related to the operator $M^\dagger M$, which is severely ill-conditioned at low 
temperatures (see {\it e.g.}~\cite{AbeSeki}). It is therefore of great interest to explore various preconditioning strategies, as these hold the key to 
finding a practical HMC implementation which is competitive with current DMC calculations.

The class of problems at hand is characterized by a zero-range interaction, such that the Hamiltonian is given by
\beq
\hat H \equiv \hat K + \hat V,
\eeq
where the kinetic energy is
\beq
\hat K = \sum_{s=\uparrow,\downarrow} \int d{\bf r} \,
\psi^\dagger_s({\bf r}) \,\frac{-\hbar^2 \nabla^2}{2m}\, \psi_s({\bf r}),
\eeq
and the potential energy
\beq
\hat V = - g \int d{\bf r} \,
\psi^\dagger_\downarrow({\bf r})\psi^{}_\downarrow({\bf r}) 
\psi^\dagger_\uparrow({\bf r})\psi^{}_\uparrow({\bf r}),
\eeq
such that $\psi^{\dagger}_s({\bf r})$ and $\psi^{}_s({\bf r})$ are creation and annihilation 
operators for particles of spin $s$ at point $\bf r$, respectively. We next seek to recast the grand canonical
partition function
\beq
\mathcal Z \equiv \text{Tr} \, \exp(-\beta (\hat H - \mu \hat N)),
\eeq
in a form amenable to a Monte Carlo calculation. We proceed by discretizing imaginary time into $N_\tau^{}$ slices of 
length $\tau = \beta/N_\tau^{}$ using a Trotter-Suzuki decomposition~\cite{Suzuki} followed by a continuous 
Hubbard-Stratonovich transformation~(HST)~\cite{HS} of the form recently proposed by Lee~\cite{Lee}. It should be noted 
that these last two steps are, to a certain extent, a matter of choice. For instance, continuous-time formulations 
exist~\cite{Rombouts} although they have as yet not been combined with HMC. The choice of HST is also dictated by 
computational preferences~\cite{Hirsch}, although it is not clear at this time whether an HMC implementation is 
compatible with a discrete HST. The end result is a path integral representation of the partition function
\beq
\label{HSZ}
\mathcal Z \, = \int \mathcal D \sigma  
\left(\det M[\sigma]\right)^2 = \int \mathcal D \sigma \, \det M^\dagger_{}[\sigma] M[\sigma],
\eeq
\beq
\label{MatrixM}
M[\sigma] \equiv 
\left( \begin{array}{ccccccc}
1 & 0 & 0 & 0 & \dots & B_{N_\tau^{}}^{}[\sigma]\\
-B_{1}^{}[\sigma] & 1 & 0 & 0 & \dots & 0 \\
0 & -B_{2}^{}[\sigma] & 1 & 0 &  \dots & 0 \\
\vdots & \vdots & \vdots & \vdots & \vdots & \vdots \\
0 & 0 & \dots & -B_{N_\tau^{} - 2}^{}[\sigma] & 1 & 0 \\
0 & 0 & \dots & 0 & -B_{N_\tau^{} - 1}^{}[\sigma] & 1
\end{array} \right),
\eeq
where the explicit form of the block matrices 
\beq
B_j^{}[\sigma] \,\equiv\, \exp(-\tau K) \, \left(1 + A \sin[\sigma_j^{}]\right),
\eeq
and $A \equiv \sqrt{2}\sqrt{\exp(\tau g) - 1}$, which results from the specific choice of HST. In momentum space, the
kinetic energy is given by
\beq
K_{{\bf k},{\bf k}'}^{} \equiv \delta_{{\bf k},{\bf k}'}^{} \, E({\bf k}),
\eeq
where we have chosen the dispersion relation $E({\bf k}) = \hbar^2 {\bf k}^2/2m$ for our calculations.
The integer vectors $({\bf k}, {\bf k}')$ assume values on a three-dimensional momentum lattice given by
$k_i^{} = 2\pi n_i^{}/N_x^{}$, where $n_i^{} = -N_x^{}/2,\ldots,N_x^{}/2-1$. Eq.~(\ref{MatrixM}) provides a 
representation of the fermionic operator $M$ referred to earlier, which is real and manifestly not symmetric.

One of the most popular and successful implementations of HMC is known as the $\varphi$-algorithm~\cite{HMC_Gottlieb}, 
which combines the stochastic evaluation of the fermion determinant in Eq.~(\ref{HSZ}) with the MD evolution of the 
auxiliary Hubbard-Stratonovich field $\sigma$ (originally the gauge field in the case of Lattice QCD). To this end, the
pseudofermion representation
\beq
\det M^\dagger_{}[\sigma] M[\sigma] \equiv \int \mathcal D \phi^\dagger \mathcal D \phi \; \exp(-S_{p}^{}[\sigma]),
\eeq
is applied, giving
\beq
S_{p}^{}[\sigma] \,\equiv \sum_{{\bf n},\tau} \left[ \phi^\dagger Q^{-1}[\sigma] \phi \right]_{{\bf n},\tau}, 
\eeq
for the corresponding action, where $Q \equiv M^\dagger_{} M$. The next step is to introduce an auxiliary field $\pi$ that will 
play the role of a conjugate momentum to the field $\sigma$ in the molecular dynamics evolution. This is accomplished 
by multiplying the partition function by a constant in the form of a path integral over $\pi$,
\beq
\mathcal Z \, \equiv  
\int \mathcal D \sigma \mathcal D \pi \mathcal D \phi^\dagger \mathcal D \phi \; 
\exp(-{\mathcal H}[\sigma,\pi]),
\eeq
where the MD Hamiltonian is given by
\beq
\mathcal H \equiv \sum_{{\bf n},\tau} \frac{\pi_{{\bf n},\tau}^2}{2} + S_{p}^{}[\sigma,\pi].
\eeq
In the pseudofermion formulation, field configurations are sampled by evolving $\sigma$ and $\pi$ according to the classical
MD equations of motion that follow from the MD Hamiltonian. It should be noted that the pseudofermion field $\phi$ 
may be sampled exactly from a Gaussian heat bath and is kept constant during the MD evolution. In practice, the MD 
evolution should be performed using a reversible symplectic integrator in order to ensure detailed balance. 
At each step in the MD evolution the ``fermion force''
\beq
F_{{\bf n},\tau}[\sigma,\phi] \equiv 
\phi^\dagger Q^{-1}[\sigma] \, \frac{\delta Q}{\delta \sigma({\bf n},\tau)} \, Q^{-1}[\sigma] \, \phi,
\eeq
is calculated, which requires frequent computation of the vector $\eta \equiv Q^{-1}_{}[\sigma] \phi$. This represents the most 
time-consuming part of the HMC algorithm. One of the most obvious advantages of the $\varphi$-algorithm is that the 
direct calculation of a large determinant is avoided, in favor of the solution of a much smaller linear problem a fixed 
number of times. Specifically, in our case this involves the repeated solution of Eq.~(\ref{MTMLinearProblem}) in its 
preconditioned form given by Eq.~(\ref{PMTMLinearProblem}), where the structure of the matrix is constant but the 
auxiliary field $\sigma$ varies at each MD step. This should be contrasted with the cost of computing the full inverse, 
which is much more expensive than solving a single linear problem. By avoiding the calculation of the determinants and 
inverses, this algorithm enables global updates of the auxiliary field $\sigma$ via MD evolution. Moreover, all HMC approaches can 
be rendered free of systematic errors associated with a finite MD integration step size by means of a Metropolis 
accept/reject step at the end of each MD trajectory.

In Sec.~\ref{Cheb}, we study the effect of preconditioning via Chebyshev polynomials, which provide an approximation to 
$Q^{-1}_{}$. Chebyshev polynomials provide a generic approach which is easy to apply, but computationally expensive.
In Sec.~\ref{WCE}, we proceed to study the weak-coupling approximation~(WCA) method of preconditioning $M$ and its
transpose separately. In Sec.~\ref{SCA} we explore the strong-coupling approximation~(SCA) to $Q^{-1}_{}$ which was 
originally proposed in Ref.~\cite{Scalettaretal}. Our findings are summarized in Sec.~\ref{Results}, followed by a 
concluding discussion in Sec.~\ref{Conclusions}.

%%%%%%%%%%%%%%%%%%%%%%%%%%%%%

\section{\label{Cheb}Chebyshev polynomials}

Chebyshev polynomials have frequently been applied in the preconditioning of relativistic quantum field theory problems. 
In fact, they have led to the development of a whole new class of HMC-type algorithms commonly referred to as
Polynomial Hybrid Monte Carlo~(PHMC)~\cite{FrezzottiJansen}, in which preconditioners are used to separate modes that
evolve at different rates in an MD trajectory. The Chebyshev preconditioning technique is based on approximating 
$z^{-1}_{}$ with a Chebyshev polynomial $P(z)$ of degree $2n$,
\beq
\label{ChebInv}
P(z) \equiv c_{2n}^{}\prod_{k=1}^{2n} (z - z_{2n,k}^{}),
\eeq
where the coefficients are given by
\beq
c^{-1}_{2n} = \frac{1+\epsilon}{2} \prod_{k=1}^{2n}\left(\frac{1+\epsilon}{2}-z_{2n,k}^{}\right),
\eeq
and the roots by
\beq
z_{2n,k}^{} = \frac{1 + \epsilon}{2}\left[1 - \cos\left(\frac{2\pi k}{2n + 1}\right) \right] -
i \sqrt{\epsilon}\sin\left(\frac{2 \pi k}{2n + 1}\right),
\eeq
such that the parameter $0 < \epsilon < 1$ provides a lower bound for the range of validity of the approximation. Indeed, 
in the interval $[\epsilon,1]$ the Chebyshev polynomials provide a good approximation, in the sense that 
the relative error
\beq
|R_{2n,\epsilon}^{}| \leq 
2\left(\frac{1-\sqrt{\epsilon}}{1+\sqrt{\epsilon}}\right)^{2n+1}_{},
\eeq
where
\beq
R_{2n,\epsilon}^{}(z) = \left[ P_{2n,\epsilon}^{}(z) - 1/z \right] z,
\eeq
is exponentially suppressed with the degree of the polynomial. 

The preconditioner we seek is obtained from Eq.~(\ref{ChebInv}) upon replacing $z \to M^\dagger_{}M$, assuming that the eigenvalue 
spectrum of ${M}^\dagger_{}{M}$ is normalized in the range $[\epsilon, 1]$. One of the most appealing features of the Chebyshev 
preconditioning method is that its speed and effectiveness can be tuned by varying the degree of the polynomial $n$. 
Moreover, since $P$ is applied in its factorized form, the speed of the preconditioner is directly linked to the speed 
with which $M$ is applied. The order in which the factors of Eq.~(\ref{ChebInv}) are applied turns out to be critical 
for numerical stability, due to cancellations between the various terms that occur naturally when evaluating a 
polynomial in finite precision arithmetic. In practice it becomes crucial beyond $n \simeq 20$ to address this issue. 
Various orderings of the factors were explored in Ref.~\cite{Bunk}, as well as a recursive Clenshaw algorithm which 
largely eliminates the accumulation of round-off error. Our experience indicates that the performance of the 
``bit-reversal'' ordering of Ref.~\cite{Bunk} is nearly identical to that of the Clenshaw algorithm, and it is slightly 
faster and requires less memory, in addition to being much simpler to implement. The bit-reversal ordering will therefore
be our method of choice.

A serious disadvantage of the Chebyshev approach is that it is only applicable to the preconditioning of the normal 
equations, as it requires the matrix to have positive definite eigenvalue spectrum. Also, as ${M}^\dagger_{}{M}$ becomes extremely 
ill-conditioned at low temperatures, polynomial orders larger than $n \sim 64$ are needed for effective preconditioning, 
which makes the Chebyshev approach rather computationally intensive. As will become evident in Sec.~\ref{Results}, this 
problem can largely be avoided by separately inverting $M$ and $M^\dagger_{}$, which however requires a different type of 
preconditioner. We shall now turn to the description of such strategies.

%%%%%%%%%%%%%%%%%%%%%%%%%%%%%

\section{\label{WCE}Weak coupling expansion}

In this section, we demonstrate how the weak-coupling limit of $M$ can be used as the starting point of an efficient 
preconditioning strategy for $M$ and $M^\dagger_{}$. It is useful to note that the representation of $M$ given in 
Eq.~(\ref{MatrixM}) can be decomposed as
\beq
M \equiv M_0^{} + A M_1^{}
\eeq
where $M_0^{} \equiv \lim_{A \to 0} M$ corresponds to the non-interacting case, and
\beq
\label{MatrixM_1}
M_1^{} \equiv 
\left( \begin{array}{ccccccc}
0 & 0 & 0 & 0 & \dots & C_{N_\tau^{}}^{}\\
-C_{1}^{} & 0 & 0 & 0 & \dots & 0 \\
0 & -C_{2}^{} & 0 & 0 &  \dots & 0 \\
\vdots & \vdots & \vdots & \vdots & \vdots & \vdots \\
0 & 0 & \dots & -C_{N_\tau^{} - 2}^{} & 0 & 0 \\
0 & 0 & \dots & 0 & -C_{N_\tau^{} - 1}^{} & 0
\end{array} \right),
\eeq
where the block matrices are given by
\beq
C_j^{} \,\equiv\, \exp(-\tau K) \, \sin[\sigma_j^{}]. 
\eeq
The inverse $M^{-1}_{}$ can then be formally expanded in powers of $AX$, giving
\beq
\label{Eq:WCE}
M^{-1}_{} = (1 + AX)^{-1}_{} M_0^{-1} = (1 - AX + A^2X^2 - \ldots + \ldots) \, M_0^{-1},
\eeq
where $X \equiv M_0^{-1} M_1^{}$. One can then determine {\it a posteriori} whether the truncated series represents a 
good approximation to $M^{-1}_{}$, which is a reasonable expectation for $A \ll 1$. The most computationally intensive 
part of such an approximation is the application of $M_0^{-1}$. However, this matrix is diagonal in frequency-momentum space. 
It is thus convenient to apply $M_0^{-1}$ after a four-dimensional Fourier transform of the relevant vector, followed by an anti-Fourier
transform to return to the original basis. This approach is referred to as Fourier acceleration. 
In frequency-momentum space, $M_0^{-1}$ is given by 
\beq
[M_0]^{-1}_{\omega, {\bf k}; \omega', {\bf k}'} \sim 
\delta_{\omega,\omega'}^{} \delta_{{\bf k},{\bf k}'}^{} \left[1 - \exp{(-i\omega -\tau \hbar^2 {\bf k}^2/2m)}\right]^{-1},
\eeq
up to a constant normalization factor, with
\beq
\omega = (2 n_\tau^{} + 1)\pi / N_\tau^{}, \quad 
k_i^{} = 2 n_i^{} \pi / N_x^{}, \quad\quad
n_\mu^{} = -N_\mu^{}/2,\ldots,N_\mu^{}/2-1.
\eeq
Care must be taken that the boundary conditions in the temporal direction are antiperiodic, as befit fermions, 
rather than periodic as commonly assumed in FFT routines. A possible complication arises due to the fact that the radius 
of convergence of Eq.~(\ref{Eq:WCE}) can be quite limited depending on the spectrum of $X$, which is unknown for all 
practical purposes. We therefore turn to techniques which may accelerate the convergence of the expansion.

%%%%%%%%%%

\subsection{Convergence acceleration methods}

Here, we will briefly review the practical aspects of the Euler and van Wijngaarden methods for 
convergence acceleration of series expansions, without concerning ourselves with the underlying theory. The interested
reader is referred to the standard literature, see {\it e.g.} Refs.~\cite{VanWijngaarden,Lauwerier,NR}. Given a 
sequence of partial sums (in this case of sign-alternating terms)
\beq
S_{0,k}^{} = \sum_{n=0}^k (-1)^n_{} a_n^{},
\eeq
where $k = 0,\ldots, k_\mathrm{max}^{}$, the Euler method consists of defining a set of new sequences given by
\beq
\label{Euler}
S_{j+1,k}^{} = p S_{j,k}^{} + (1-p) S_{j,k+1}^{},
\eeq
where $k = 0,\ldots,k_\mathrm{max}^{}\!-\! (j\!+\!1)$. According to this method, instead of using 
$S_{0,k_\mathrm{max}^{}}^{}$ as an estimate of 
the infinite sum, one can arrive at a much better estimate as follows. Use the original sequence to define new ones 
using Eq.~(\ref{Euler}) until all the terms available have been exhausted. The last sum, namely 
$S_{k_\mathrm{max}^{},0}^{}$ will 
consist of only one term, which is the estimate we seek. In practice one usually takes $p=1/2$, and it is not uncommon 
to achieve convergence improvements by several orders of magnitude with a handful of terms as a starting point. It has 
also been pointed out in the literature that $S_{2 k_\mathrm{max}^{}/3, k_\mathrm{max}^{}/3}$ is often an even better 
approximation than $S_{k_\mathrm{max}^{},0}^{}$. We shall apply this method to the series that results of applying 
Eq.~(\ref{Eq:WCE}) to a given vector.

The sequence $S_{j,0}^{}$ is commonly referred to as the Euler transform of the original sequence 
$S_{0,k}^{}$, and it can be shown to be given by
\beq
S_{j,0}^{} = \sum_{n=0}^{j} b_n^{}
\eeq
where
\beq
b_n^{} \equiv \sum_{j=0}^n \left(
\begin{array}{c}
\!n\!\\
\!j\!
\end{array}
\right)
 p^{n-j}_{} (1-p)^{j+1}_{} (-1)^j_{} a_j^{}.
\eeq
The idea of van Wijngaarden was to first multiply each term in the original series by 
non-vanishing arbitrary constants $\lambda_k$ and then perform an Euler transformation. It was assumed that a 
moment generating function $\phi(t)$ exists such that
\beq
\lambda_k^{-1} = \int_0^{\infty} \phi(t) \, t^k dt,
\eeq
and it was shown that the sum of the original series is given by
\beq
\sum_{n=0}^\infty (-1)^n_{} a_n^{} = \sum_{n=0}^\infty \mu_n^{} b_n^{},
\eeq
where
\beq
\mu_k^{} \equiv \int_0^{\infty} dt \, \phi(t) \, \frac{t^k_{}}{(1-p + pt)^{k+1}_{}}.
\eeq
In practice, van Wijngaarden's transformation can turn a slowly convergent or even divergent series into a rapidly 
convergent series. It can also be shown that the so-called special van Wijngaarden transformation, for 
which $\lambda_k = s^k/k!$ and $\phi(t) = s \exp(-s t)$, where $s$ is a free parameter, does not change the Borel sum 
of the original series and corresponds to the Laplace transform of the Euler transformation~\cite{Lauwerier}.

%%%%%%%%%%%%%%%%%%%%%%%%%%%%%

\section{\label{SCA}Strong coupling approximation}

The idea of constructing a preconditioner based on a strong-coupling approximation was originally developed by Scalettar 
{\textit et al.} in Ref.~\cite{Scalettaretal}. It consists of constructing the 
inverse of a strong-coupling approximation to $M^\dagger_{}M$, which we shall denote by $\tilde{M}^\dagger_{} \tilde{M}$, 
where $\tilde M$ has the same structure as $M$ but with the substitution
\beq
B_j^{} \to B_{0,j}^{}  = 1 + A \sin[\sigma_j^{}],
\eeq
where it should be noted that $B_{0,j}^{}$ is diagonal in coordinate space. In this approach, all the blocks will have 
this property, which makes it extremely inexpensive from a computational point of view. In order to find the inverse, 
one defines a factorization
\beq
\tilde{M}^\dagger_{} \tilde{M} \equiv L^\dagger D L,
\eeq
where
\beq
L^\dagger_{} = 
\left( \begin{array}{ccccccc}
1 & 0 & 0 & 0 & \dots & 0\\
-L_{1}^{} & 1 & 0 & 0 & \dots & 0 \\
0 & -L_{2}^{} & 1 & 0 &  \dots & 0 \\
\vdots & \vdots & \vdots & \ddots & \vdots & \vdots \\
0 & 0 & \dots & -L_{N_\tau^{} - 2}^{} & 1 & 0 \\
L_{N_\tau} & 0 & \dots & 0 & -L_{N_\tau^{} - 1}^{} & 1
\end{array} \right), \quad\quad
\eeq
and
\beq
D = 
\left( \begin{array}{ccccccc}
D_1^{} & 0 & 0 & 0 & \dots & 0\\
0 & D_2^{} & 0 & 0 & \dots & 0 \\
0 & 0 & D_3^{} & 0 &  \dots & 0 \\
\vdots & \vdots & \vdots & \ddots & \vdots & \vdots \\
0 & 0 & \dots & 0 & D_{N_\tau^{} - 1}^{} & 0 \\
0 & 0 & \dots & 0 & 0 & D_{N_\tau^{}}^{}
\end{array} \right),
\eeq
for which it is straightforward to show that the individual blocks are given by
\bea
L_t^{} &=& D_t^{-1} B_{0,t}^{}, \\
D_t^{} &=& \alpha + B^2_{0,t} - B_{0,t-1}^{} D_{t-1}^{-1} B_{0,t-1}^{},
\eea
for $t=1, \ldots , N_\tau^{}$ 
and
\bea
L_{N_\tau^{}}^{} &=& D_1^{-1} B_{0,N_\tau^{}}^{}, \\
D_{N_\tau^{}}^{} &=& \alpha + B^2_{0,{N_\tau^{}}} - 
B_{0,{N_\tau^{}}^{}-1}^{} D_{{N_\tau}-1}^{-1} B_{0,{N_\tau^{}}-1}^{} - B_{0,{N_\tau^{}}}^{} D_{1}^{-1} 
B_{0,{N_\tau^{}}}^{},
\eea
where the parameter $\alpha$ is tuned such that the number of CG iterations is minimized.
The matrix
\bea
L^{-1}_{} D^{-1}_{} L^{\dagger-1}_{} = (\tilde{M}^\dagger_{} \tilde{M})^{-1}_{} 
\eea
then plays the role of the preconditioner in this approach. Notice that the matrices $L$ and $L^\dagger_{}$ are trivial to 
invert, in the sense that the linear problems $Lx = y$ and $L^\dagger x = y$ can be solved in a fast and straightforward 
fashion. For example, $X = L^{\dagger -1}_{} Y$ is given by the recursive expression
\bea
x(r,1) &=& y(r,1), \nonumber \\
x(r,2) &=& y(r,2) +  L_1^{} \,x(r,1), \nonumber \\
x(r,3) &=& y(r,3) +  L_2^{} \,x(r,2), \nonumber \\
&\ldots& \nonumber \\
x(r,{N_\tau}) &=& y(r,{N_\tau}) +  L_{N_\tau-1}^{} \, x(r,{N_\tau}-1) - L_{N_\tau} \, x(r,1).
\eea
While Ref.~\cite{Scalettaretal} found this approach promising, it presents at least two disadvantages. Firstly, 
it is not easy to improve in a systematic fashion, in contrast with the Chebyshev approximation. Indeed, one of the 
key properties of this preconditioner is that it only involves diagonal matrices. Any improvement involving 
the kinetic energy operator will eliminate this property and make the approach much more computationally expensive. 
Secondly, just like the Chebyshev method, this preconditioner aims at approximating the inverse of ${M}^\dagger_{}{M}$, whereas 
inverting $M$ and $M^\dagger_{}$ separately is preferable due to the dramatic increase in the condition number of ${M}^\dagger_{}{M}$ relative 
to $M$. However, it should be pointed out that this preconditioner is computationally very inexpensive and can 
furthermore be combined with the Chebyshev preconditioner, using the latter as a secondary filter. It is not yet known 
whether such a combined approach is useful in practice.

%%%%%%%%%%%%%%%%%%%%%%%%%%%%%

\section{\label{Results}Results}

\begin{figure}[t]
\centering
\epsfig{file=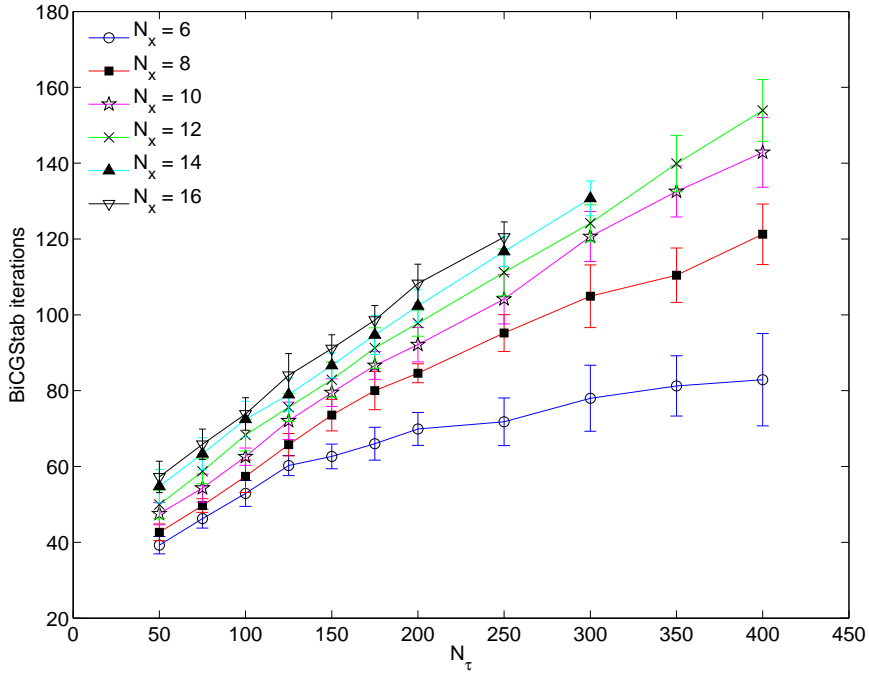, width=.8\columnwidth}
\caption{BiCGStab iterations as a function of $N_\tau^{}$ and $N_x^{}$, for the solution of $M^\dagger_{} y = b$ followed by 
$M x = y$, for a tolerance of $10^{-7}_{}$ and a coupling of $g = 5.0$ $(A \simeq 0.75)$, using the second-order 
weak-coupling expansion~(\textit{WCE2}) as preconditioner. The results represent averages over~20 random auxiliary field 
configurations.
\label{Fig:iter}}
\end{figure}

The number of CG or BiCGStab iterations required to solve the linear problems of Eqs.~(\ref{MTMLinearProblem}) and
~(\ref{MLinearProblem}) is given in Table~\ref{TableIterations} for the various preconditioners, with a more comprehensive 
study of the scaling of the second-order weak-coupling preconditioner~(WCE2) in Fig.~\ref{Fig:iter}. The convergence criterion, referred to as the 
tolerance parameter, was taken to be a reduction in the norm of the residual by a factor of $10^{-12}$. The achieved 
absolute accuracy is approximately constant for each space-time volume $N_x^3\times N_\tau^{}$, but deteriorates slowly 
as $N_x^{}$ and $N_\tau^{}$ are increased. However, it should be noted that in typical HMC simulations, a tolerance 
parameter of $\sim 10^{-7}$ is considered adequate. Our test runs have been performed using a imaginary time step of 
$\tau = 0.05$, with couplings of $g = 2.5, 5.0$ and~$7.5$. The corresponding values of $A$ are $\simeq 0.52, 0.75$ 
and~$0.95$. The results shown in Tables~\ref{TableIterations} and~\ref{TableMVs} correspond to $g = 5.0$, which is close 
to the unitary limit.

We have used auxiliary field configurations $\sigma$ with randomly distributed entries in the interval $[-\pi,\pi]$ in 
order to characterize the performance of the preconditioners. While this is the proper interval for the chosen HST,
the distribution differs from that of a thermalized HMC simulation, where the values of $\sigma$ tend to cluster around
$-\pi/2$ and $\pi/2$. However, in practice the net effect of using thermalized instead of random configurations is to 
increase the number of CG iterations by a factor of $\sim 2$ across all parameter values. We have confirmed this 
behavior in a number of cases, and we therefore conclude that the relative merits of the preconditioners remain 
unaffected by the issue of thermalized versus random configurations. We have also found that~10 random field 
configurations suffices to provide an estimate of the average number of CG iterations with an uncertainty of less than
$10-15\%$ in all cases. The uncertainly can be further suppressed by increasing the number of sample 
configurations. So far, we have made limited use of Euler--Van Wijngaarden acceleration techniques. 
For the second-order weak-coupling expansion~(WCE2), the simplest Euler method yielded a moderate but definite 
improvement over the non-accelerated implementation. For the fourth-order expansion (WCE4), the Euler method was found 
to reduce the number of iterations by up to a factor of 2. Finally, we have varied the right-hand side from a 
constant vector to one with randomly distributed entries and found no impact on the results. 

The number of MV operations consumed by each method is shown in Table~\ref{TableMVs}. Per iteration, CG 
on the normal equations requires one application of $M^\dagger_{}M$ as well as the preconditioner $P(M^\dagger_{}M)$. On the other hand, BiCG 
and BiCGStab require the application of both $M$ and $M^\dagger_{}$ (as well as their respective preconditioners) for each 
iteration. Thus, in the absence of a preconditioner, the cost per iteration of CG, BiCG and BiGStab is the same. The 
main advantage of using the normal equations with CG is that one avoids the need to successively solve two linear 
problems. The MV operations performed per CG iteration when using Chebyshev preconditioning can be estimated as follows: 
the application of a polynomial of degree $d$ to a vector requires $2d$ MV operations; adding the application of the 
matrix $M^\dagger_{}M$ itself yields a total of $2(d + 1)$ operations. Estimating the computational cost of the weak-coupling 
preconditioner is somewhat more involved. The BiCG or BiCGStab iterations contribute two MV operations from 
the application of $M^\dagger_{}$ and $M$. For each power of $A$, applying $K$ contributes one MV operation to the total cost. 
However, the non-interacting matrix $M_0^{}$ is applied using a four-dimensional FFT, which scales as $N_x^3 \times 
N_\tau^{} \times \log(N_x^3 \times N_\tau^{})$, which differs from the $N_x^3 \times N_\tau^{} \times \log(N_x^3)$ 
scaling of the MV operations. The overall computational cost of the approach using the 
weak-coupling preconditioner of degree $d$ in $A$ is given by $2(d + 1)(1 + 2\beta)$ MV operations, where $\beta$ is
the factor in front of the scaling law for the 4D FFT, relative to that of the 3D FFT. While we have not attempted to estimate $\beta$ directly
(it is implementation-dependent), we find that in practice the gain in CPU time can be lower by a factor of $\sim 2$ compared with the 
gain factors quoted in Table~\ref{TableMVs}, where we set $\beta = 1$. Notice however that the gain is still substantial, especially 
as the total number of MV operations increases only mildly as a function of $N_\tau^{}$. 

\begin{table}[t]
\begin{center}
\caption{\label{TableIterations}
Summary of the number of iterations required for the solution of $M^\dagger_{}M x = b$ with CG, or $M^\dagger_{} y = b$ followed by $M x = 
y$ with BiCGStab, for a tolerance of $10^{-12}_{}$ and a coupling of $g = 5.0$ $(A \simeq 0.75)$, which is close to the 
unitary limit. The columns labeled ``\textit{Cheb8}'', ``\textit{Cheb16}'' and ``\textit{Cheb32}'' denote, respectively, 
Chebyshev preconditioning of degrees 8, 16 and~32 using bit-reversal ordering of the roots. The strong-coupling 
preconditioner is labeled ``\textit{SCA}'', and ``\textit{WCE0}'', ``\textit{WCE1}'' and 
``\textit{WCE2}'' denote the weak-coupling expansion of order 0, 1 and~2 in $A$, respectively. The 
reported numbers represent averaged over~10 random auxiliary field configurations, which yields an uncertainty 
below~10-15\%. A line denotes the cases that failed to converge in less than 10,000 iterations.} 
\vspace{.5cm}
\begin{tabularx}{\textwidth}{@{\extracolsep{\fill}}c c c c c c c c}
	\hline {Lattice size} & \multicolumn{4}{c}{CG} & \multicolumn{3}{c}{BiCGStab} \\ 
	\cline{2-5} \cline{6-8}
	$N_x^3\times N_\tau^{}$ & \textit{Cheb8} & \textit{Cheb16} & \textit{Cheb32} & \textit{SCA} & \textit{WCE0} & \textit{WCE1} & \textit{WCE2} \\  
	\hline \hline
	$6^3 \times 50$ 			&  294		& 167		& 104		& 1,790	 	& 128 	& 77		& 55	\\  
	$8^3 \times 50$ 			&  386 		& 226		& 130		& 2,388		& 147	& 84		& 62	\\  
	$10^3 \times 50$			&  489 		& 259		& 163		& 2,810	 	& 152 	& 93		& 67	\\  
%%%% 
	\hline
	$6^3 \times 100$ 			&  534 		& 316		& 187		& 3,470		& 162 	& 97		& 71	\\  
	$8^3 \times 100$ 			&  875		& 460		& 288		& 5,223		& 183 	& 114	& 78	\\  
	$10^3 \times 100$ 			&  1,102 		& 610		& 382		& 7,104 		& 196 	& 118	& 88	\\  
%%%%
	\hline
	$6^3 \times 200$ 			&  864		& 530		& 276		& 5,987		& 208	& 119	& 87	\\  
	$8^3 \times 200$ 			&  3,301 		& 998		& 604 		& ---  		& 246 	& 148	& 105\\  
	$10^3 \times 200$ 			&   ---		& 1,306		& 735 		& --- 			& 279 	& 162	& 121 \\
\hline
\end{tabularx}
	
\end{center}
\end{table}

While the strong-coupling approximation is extremely inexpensive from a computational point of view, its effect on the 
solution process is rather mild, at least for the values of $g$ considered here. Thus, the number of CG 
iterations remains rather high, which is an undesirable feature as the iterative solution process is prone to 
accumulate numerical round-off error after seveal hundred iterations, which is especially problematic for the BiCG and 
BiCGStab solvers. As a rule of thumb, in our HMC calculations we aim to keep the iterations below the 
$200\!-\!300$ range. Though computationally cheap, a preconditioner which is unable to reduce the iteration count below 
$\sim 1000$ is of limited practical usefulness, especially at large $N_\tau^{}$ where the number of iterations may 
increase dramatically. The effective cost of the strong-coupling preconditioner in terms of MV operations 
is difficult to estimate, as it involves no FFTs. We have found, empirically, that the number of MV operations required 
per CG iteration can be conservatively estimated to be $\sim 3$. In light of the exceedingly high iteration 
count shown in Table~\ref{TableIterations} for this preconditioner, however, this MV estimate is irrelevant.

\begin{table}[t]
\begin{center}
\caption{\label{TableMVs}
Summary of the number of matrix-vector (MV) operations required for the solution of $M^\dagger_{}M x = b$ with CG, or $M^\dagger_{} y = b$ 
followed by $M x = y$ with BiCGStab. Notation and parameters are as for Table~\ref{TableIterations}. The rightmost 
column illustrates the gain provided by the lowest-order weak-coupling expansion using BiCGStab over the most favorable
case using CG. As explained in the text, the actual gain in CPU time is likely to be somewhat less. Notice in particular 
the mild increase in the number of MV operations for BiCGStab as a function of $N_x^{}$ and $N_\tau^{}$.}
\vspace{.5cm}
	\begin{tabularx}{\textwidth}{@{\extracolsep{\fill}}c c c c c c c c c}	
	\hline \multicolumn{1}{c}{Lattice size} & \multicolumn{4}{c}{CG} & \multicolumn{3}{c}{BiCGStab} & \multicolumn{1}{c}{} \\
	\cline{2-5} \cline{6-8}
	$N_x^3\times N_\tau^{}$ & \textit{Cheb8} & \textit{Cheb16} & \textit{Cheb32} & \textit{SCA} & \textit{WCE0} & \textit{WCE1} & \textit{WCE2} & {Gain} \\
	\hline\hline
	$6^3 \times 50$ 		& 5,292	& 5,678	& 6,864	& 5,370	& 768	& 924	& 990 & {\bf 7}x \\  
	$8^3 \times 50$ 		& 6,948 	& 7,684	& 8,580	& 7,164	& 882	& 1,008	& 1,116	& {\bf 8}x \\  
	$10^3 \times 50$ 		& 8,802 	& 8,806 	& 10,758 	& 8,430 	& 912	& 1,116	& 1,206	& {\bf 9}x \\  
%%%% 
\hline
	$6^3 \times 100$ & 9,612 & 10,744 & 12,342 & 10,410 & 972 & 1,164 & 1,278 & {\bf 10}x \\  
	$8^3 \times 100$ & 15,750 & 15,640 & 19,008 & 15,669 & 1,098 & 1,368 & 1,404 & {\bf 14}x \\  
	$10^3 \times 100$ & 19,836 & 20,740 & 25,212 & 21,313 & 1,176 & 1,416 & 1,584 & {\bf 17}x \\  
%%%%
\hline
	$6^3 \times 200$ & 15,552 & 18,020 & 18,216 & 17,961 & 1,248 & 1,428 & 1,566 & {\bf 12}x \\  
	$8^3 \times 200$ & 59,418 & 33,932 & 39,864 & ---    & 1,476 & 1,776 & 1,890 & {\bf 23}x \\
	$10^3 \times 200$ &  --- & 44,404 & 48,510 & --- & 1,674 & 1,944 & 2,178 & {\bf 27}x \\
\hline
	\end{tabularx}
\end{center}
\end{table}

We have also tested the weak-coupling approximation with CG and found that it provides little benefit, in 
contrast to the dramatic improvement when used with BiCGStab. This situation arises as we have only attempted left 
preconditioning with CG, and thus the full potential of the preconditioning strategy is not realized. For this to be the 
case, one would need to precondition with $(M_0^\dagger)^{-1}$ on the left and with $M_0^{-1}$ on the right. In the case of BiCGStab 
this complication does not appear, as $M$ and $M^\dagger_{}$ are preconditioned separately. Finally, in Table~\ref{TableUs} we 
show the behavior of the weak-coupling preconditioner as a function of the coupling, for 
$g = 2.5, 5.0$ and $7.5$. We note that the case of $g = 10.0$ ($A \simeq 1.14$), typically failed to converge 
in less than 15,000~MV operations.

\begin{table}
\caption{\label{TableUs}
Effectiveness of the second-order~(\textit{WCE2}) and fourth-order~(\textit{WCE4}) 
weak-coupling preconditioners as a function of $g$. For $g = 2.5$, we have $A \simeq 0.52$, for $g = 5.0$, $A \simeq 0.75$ 
and for $g = 7.5$, $A \simeq 0.95$. The figures given represent the number of matrix-vector operations required to solve 
$M^\dagger_{} y = b$ followed by $M x = y$ with BiCGStab. These are averages over 10 random configurations, with an 
uncertainly in the range of 10-15\%.}
\begin{center}
\begin{tabularx}{\textwidth}{@{\extracolsep{\fill}}c c  c  c  c  c  c}
	\hline {Lattice size} & \multicolumn{2}{ c }{$g = 2.5$} & \multicolumn{2}{c}{$g = 5$} & \multicolumn{2}{c}{$g = 7.5$} \\
	\cline{2-3} \cline{4-5} \cline{6-7}
	$N_x^3\times N_\tau^{}$ & \textit{WCE2}	& \textit{WCE4}	& \textit{WCE2}	& \textit{WCE4}	& \textit{WCE2} & \textit{WCE4} \\
	\hline\hline 
	$10^3 \times 50$ 		& 630		& 600		& 1,206		& 1,320		& 2,358		& 4,110	\\ 
	$10^3 \times 100$ 		& 702		& 690		& 1,584		& 1,530		& 3,816		& 9,330	\\ 
	$10^3 \times 200$ 		& 864		& 780 		& 2,178		& 2,190		& 6,444		& 12,720  \\
\hline
\end{tabularx}
\end{center}
\end{table}

%%%%%%%%%%%%%%%%%%%%%%%%%%%%%

\section{\label{Conclusions}Summary and Conclusions}

In this work, we have evaluated three different strategies to precondition the non-relativistic many-fermion problem: 
Chebyshev polynomials, a strong-coupling approximation and a weak-coupling expansion. We have argued that such 
preconditioning forms a central part of HMC calculations, as the frequent solution of a linear 
problem involving the ill-conditioned fermion matrix $M$ is at the heart of the HMC algorithm. We have considered the 
cases of normal equations $M^\dagger_{}M x = b$ which is tractable using the CG algorithm, as well as the 
two-step approach of solving $M^\dagger_{} y = b$ followed by $M x = y$, using the BiCG or BiCGStab algorithms. Our 
results indicate that both the Chebyshev polynomials (Sec.~\ref{Cheb}) and the weak-coupling expansion (Sec.~\ref{WCE}) 
can be effective preconditioners, especially when high orders are used. However, from the point of view of performance, 
Chebyshev polynomials represent an expensive choice. Additionally, the Chebyshev approach requires a matrix with 
a positive definite eigenvalue spectrum, which forces us to work with the CG solver on the normal equations. This is 
unfortunate, as $M$ tends to be a rather ill-conditioned matrix (especially at low temperatures), which 
makes the normal system $Q x = b$ extremely ill-conditioned. The result is that the number of MV operations required 
to solve the problem with a preset accuracy grows rapidly with the size of the problem, in particular with $N_\tau^{}$, 
which controls the temperature.

We have shown that solving $M^\dagger_{} y = b$ followed by $M x = y$ using the BiCGStab algorithm provides an alternative, 
extremely promising approach. The use of a weak-coupling preconditioner supplemented with a simple convergence 
acceleration technique solves the problem elegantly and provides dramatically enhanced performance in terms of both 
number of iterations and MV operations, which translates into significant savings of CPU time and improved scaling of 
the HMC algorithm at low temperature. While we have only provided a first, sketchy comparison of the two methods, 
the difference between using CG with Chebyshev preconditioning and BiCGStab with weak-coupling preconditioning is so 
striking that the latter technique is the obvious choice. The weak-coupling preconditioner represents a good example in 
which knowledge of the structure of the matrix, as well as the physical problem at hand, allows for the construction of 
an effective and efficient strategy to accelerate the solution process.

The strong-coupling approximation of Ref.~\cite{Scalettaretal} provides an alternative to Chebyshev preconditioning.
While computationally inexpensive, this preconditioner turned out to be less efficient in terms of its ability to reduce 
the number of CG iterations. Apparently, this approach fails to capture the relevant physics at the range of couplings 
in the vicinity of the unitary limit. In an effort to shed more light on this problem, we explored the behavior of the 
weak-coupling preconditioner as a function of the coupling $g$, and found that its effectiveness breaks down somewhere 
between $g = 7.5$ and~$10$. This situation might be remedied by using a strong-coupling expansion in powers of 
$A^{-1}_{}$, although our attempts to apply such an expansion have so far not been successful. However, we believe this 
to be a topic worthy of further investigation. Alternatively, one can reduce the value of the imaginary time step 
$\tau$, which serves to extend the usefulness of the weak-coupling expansion to larger values of $g$, given that the 
relevant expansion parameter is $A = \sqrt{2}\sqrt{\exp(\tau g) - 1}$. However, this has the drawback of 
increasing the value of $N_\tau^{}$ required to achieve a given temperature, which can be taxing on the method.

Various cases of interest that could and should be studied lie beyond the scope of this work. Among these are the case 
of finite effective range, relevant for nuclear and neutron matter calculations, which would make the potential energy 
operator more dense in real space (it is diagonal for the zero-range interaction considered here). Other dispersion 
relations, such as that of the conventional Hubbard model should be studied as well, particularly since this might 
open the door to large-scale HMC simulations in the fields of solid-state and atomic physics. Finally, we have focused 
here on a problem in $3+1$ dimensions, although the methods employed in this study carry over directly to applications 
in lower dimensions.

%%%%%%%%%%%%%%%%%%%%%%%%%%%%%

\section*{Acknowledgements}

We would like to thank Dick Furnstahl, Richard Scalettar and Nandini Trivedi for useful discussions and encouragement. 
We acknowledge support under U.S. DOE Grants No.~DE-FG02-00ER41132 and DE-AC02-05CH11231, UNEDF SciDAC 
Collaboration Grant No.~DE-FC02-07ER41457 and NSF Grant No.~PHY--0653312. 
This study was supported in part by the Academy of Finland through its Centers of Excellence Program (2006 - 2011), 
the Vilho, Yrj\"o, and Kalle V\"ais\"al\"a Foundation of the Finnish Academy of Science and Letters, and the Waldemar von Frenckell 
and Magnus Ehrnrooth Foundations of the Finnish Society of Sciences and Letters. Part of this work was performed 
using an allocation of computing time from the Ohio Supercomputer Center.

%%%%%%%%%%%%%%%%%%%%%%%%%%%%%

%\section*{References}

%%%%%%%%%%%%%%%%%%%%%%%%%%%%%

\end{document}